\documentstyle[preprint,eqsecnum,aps]{revtex}

\begin{document}
\draft \preprint{HEP/123-qed}
\title{Dimensionality Effects on the Su-Schrieffer-Heeger Model}
\author{Marco Zoli}
\address{Istituto Nazionale Fisica della Materia - Universit\'a di Camerino, \\
62032 Camerino, Italy. e-mail: zoli.marco@libero.it }

\date{\today}
\maketitle
\begin{abstract}
A low order diagrammatic study of the dimension dependent
Su-Schrieffer-Heeger model Hamiltonian in the weak electron-phonon
coupling regime is presented. Exact computation of both the charge
carrier effective mass and the electron spectral function shows
that electrons are good quasiparticles in the antiadiabatic limit
but new features emerge in the adiabatic and intermediate regime,
where the phonons and the electrons compare on the energy scale.
Here we find: i) a sizeable mass enhancement over the bare band
value, ii) the appearance of many transition peaks in the band
bottom spectral function together with a growing loss of spectral
weight at larger {\it e-ph} couplings. The onset of a polaronic
state is favoured in two dimensions.

{\it keywords: Adiabatic Polarons, Effective Mass, Electron-Phonon
Coupling}
\end{abstract}

\widetext
\section*{1.Introduction}

A sizeable electron-phonon interaction can induce a local
deformation in the lattice structure accompanied by the formation
of a quasiparticle with multiphononic character, the polaron
\cite{landau}. A wide literature has been produced on the subject
during the last decades
\cite{toyozawa,eagles,emin,devreese,romero,mello,fehske} and
particular emphasis has been recently laid on the polaronic
properties of high $T_c$ superconductors \cite{mott}. As the
spatial extension of the lattice deformation can vary, the
concepts of large and small polaron have been introduced: the
transition between a large and a small polaron state is driven by
the strength of the electron-phonon coupling
\cite{rashba,raedt,kopida,tsiro} and monitored through the
behavior of ground state properties such as the polaron energy
band and the effective mass \cite{romero1,kornilo,jeckel}. This
transition \cite{gerlow}, leading to a self trapped state
\cite{yonemitsu} at strong couplings, is accompanied by a sizeable
enhancement of the effective mass whose value  may change
considerably according to the degree of adiabaticity and the
peculiarities of the lattice structure \cite{alexkab,io2}.
Theoretical investigations usually start from the Holstein
molecular crystal model \cite{holstein} which assumes a momentum
independent coupling of electrons to dispersive optical phonons.
The coupling to acoustical phonons, although possible in principle
\cite{io2}, would lead to huge mass renormalizations
\cite{farias}. In fact, also the Holstein optical polaron masses
are very heavy (at least larger than $10^3$ times the bare band
mass) in the self-trapped state but masses of order $10$ times the
bare band mass are possible in the presence of high energy phonon
spectra \cite{io3}. As a remarkable feature of the Holstein model
the polaron mass turns out to be essentially dimension independent
for any value of the adiabaticity parameter \cite{io4}.

In some systems however \cite{barisic,schulz,lu} the {\it e-ph}
interaction modifies the electron hopping matrix elements thus
leading to a momentum dependent coupling function. In this case an
appropriate theoretical framework is offered by the
Su-Schrieffer-Heeger(SSH) model Hamiltonian \cite{ssh} originally
proposed to explain the conducting properties of quasi one
dimensional polymers as polyacetylene \cite{ssh1}. In these
systems the CH monomers form chains of alternating double and
single bonds leading to two regions with different structural
patterns having the same energy. This twofold degenerate ground
state sustains a nonlinear localized excitation, a domain wall
separating the two regions. Being the wall thickness  much larger
than the lattice constant in the weak coupling regime a continuum
version \cite{takayama} of the SSH model has been formulated thus
leading to an analytical description of the solitonic excitation
in the dimerized system \cite{hirsch}. The continuum model admits
also polaron-like solutions \cite{campbell,onodera,boyanovsky} and
a periodic array of solitons solution, the soliton lattice
\cite{horowitz,kivelson}.

In general, the SSH Hamiltonian provides an alternative (to the
Holstein model) tool to analyse the physics of polaron formation
as tuned by the strength of the {\it e-ph} coupling. In two
dimensions, the SSH model has also been studied as a particular
case of the 2D half filled Hubbard model (with zero on site
repulsion) \cite{tang} soon after the discovery of high $T_c$
superconductivity. In particular, a recent investigation
\cite{ono1} has revealed the complexity of the square lattice
structure described by a SSH type Hamiltonian pointing out that
the opening of the gap at the Fermi level (due to the Peierls
distortion) involves many lattice modes with wave numbers parallel
to the nesting vector. The static and dynamical polaronic
properties have been also analysed \cite{ono2} in the same model
and estimates of the effective mass have suggested that 2D
polarons are heavier than 1D polarons.

This paper deals with the two dimensional electron-lattice system
treating the SSH tight binding Hamiltonian by a weak coupling
perturbative method. This approach, although not adequate to
capture the full multiphononic nature of the polaronic
quasiparticle, still  can provide useful informations
\cite{khomyakov} regarding the onset of polaron formation in some
portions of parameter space. Here we look first at the mass
renormalization in one and two dimensions exploring a wide range
of values for the adiabatic parameter and, successively, we
compute the electronic spectral function to detect whether and to
which extent bare electrons behave as good quasiparticles. The
Section 2 outlines the SSH model and contains the results of this
study while some conclusions are drawn in Section 3.

\section*{2.Model and Results}

In real space the SSH Hamiltonian reads

\begin{eqnarray}
H=\,& & \sum_{{\bf r,s}}J_{{\bf r,r+s}} \bigl(f^{\dag}_{\bf r}
f_{\bf r+s} + f^{\dag}_{\bf r+s} f_{\bf r} \bigr) + \sum_{\bf r}
\Bigl({{p^2_{\bf r}}\over {2M}} +  \sum_{\bf r,s}{K \over
2}(u_{\bf r} - u_{\bf r+s})^2
 \Bigr)\,
 \nonumber \\
& &J_{\bf r,r+s}=\, - {1 \over 2}\bigl[ J + \alpha (u_{\bf r} -
u_{\bf r+s})\bigr] \label{1}
\end{eqnarray}

where the double summation over {\bf r} and {\bf s} runs over
first neighbors lattice sites. $J$ is the nearest neighbors
hopping integral and isotropic conditions are assumed in the
square lattice. $\alpha$ is the electron-phonon coupling, $u_{\bf
r}$ is the dimerization coordinate which specifies the
displacement of the ${\bf r}-$ lattice site from the equilibrium
position, $p_{\bf r}$ is the momentum operator conjugate to
$u_{\bf r}$, $M$ is the ion (ionic group) mass, $K$ is the
effective spring constant, $f^{\dag}_{\bf r}$ and $f_{\bf r}$
create and destroy electrons on the ${\bf r}-$ site. Let's expand
the lattice displacement and its conjugate momentum in terms of
the phonon creation and annihilation operators $b^{\dag}_{\bf q}$
and $b_{\bf q}$ and Fourier transform the electron operators

\begin{eqnarray}
u_{\bf r}=\,& &\sum_{\bf q} {1 \over {\sqrt{2MN \omega_{\bf q}}}}
\bigl(b^{\dag}_{\bf -q} + b_{\bf q} \bigr)\exp(i{\bf q \cdot r})\,
 \nonumber \\
p_{\bf r}=\,& & i \sum_{\bf q} \sqrt{{M \omega_{\bf q}} \over
{2N}} \bigl(b^{\dag}_{\bf -q} - b_{\bf q} \bigr)\exp(i{\bf q \cdot
r})\,
 \nonumber \\
f_{\bf r}=\,& & {1 \over {\sqrt{N}}}\sum_{\bf k} \exp(i{\bf k
\cdot r})f_{\bf k}
 \label{2}
\end{eqnarray}

in order to obtain the SSH Hamiltonian in momentum space:

\begin{eqnarray}
& &H=\, H_0 + H_{int}\,
 \nonumber \\
& &H_0=\, \sum_{\bf k}\varepsilon_{\bf k} f^{\dag}_{\bf k} f_{\bf
k} + \sum_{\bf q} \omega_{\bf q} b^{\dag}_{\bf q} b_{\bf q} \,
 \nonumber \\
& &H_{int}=\, \sum_{\bf k,q} g({\bf k+q,k}) \bigl(b^{\dag}_{\bf
-q} + b_{\bf q} \bigr) f^{\dag}_{\bf k+q} f_{\bf k} \,
 \nonumber \\
& & \varepsilon_{\bf k}=\, -J \cos ({\bf k \cdot a}) \,
 \nonumber \\
& & \omega^2_{\bf q}=\, 4{{K \over M}}\sin^2 \biggl({{\bf q \cdot
a} \over 2}\biggr) \,
 \nonumber \\
& &g({\bf k+q,k})=\, {{i \alpha} \over {\sqrt {2MN \omega_{\bf
q}}}} \bigl(\sin \bigl({\bf (k+q) \cdot a}\bigr) - \sin {\bf k
\cdot a} \bigr) \label{3}
\end{eqnarray}

where $N$ is the total number of lattice sites. $a = |{\bf a}|$ is
the lattice constant and the 2D reduced Brillouin zone is spanned
by the vectors ${\bf b}_1 = (2 \pi /a, 0)$ and ${\bf b}_2 = (0, 2
\pi /a)$. Being ${\bf q}=\,(q_x, q_y)$, the numerical integration
constraint is defined by $q_x \in [0,2\pi/a]$, $q_y \in [0,2\pi/a
- q_x]$. In 1D, the phonon dispersion relation is defined in the
range $q \in [0,\pi]$ and, in the reduced Brillouin zone, the
spectrum displays both an acoustic and an optical branch
\cite{naka}. The model contains three free parameters: the hopping
integral $J$, the zone boundary frequency
$\omega_{\pi}=\,2\sqrt{{K/M}}$ which coincides with the zone
center  optical frequency in the reduced zone scheme, the coupling
constant $\alpha^2/4K$.

The full electron propagator in the Matsubara Green's functions
formalism is defined as:

\begin{equation}
G({\bf k},\tau)=\,-\sum_{n=0}^\infty (-1)^n \int_0^\beta d\tau_1
...d\tau_n \Bigl < T_\tau f_{\bf k}(\tau) H_{int}(\tau_1) \cdot
\cdot H_{int}(\tau_n) f^\dag_{\bf k}(0) \Bigr >_0 \label{4}
\end{equation}

where $\beta$ is the inverse temperature, $T_\tau$ is the time
ordering operator, $<...>_0$ indicates that thermodynamic averages
are taken with respect to the unperturbed Hamiltonian and only
different connected diagrams contribute to any order $n$. I have
computed exactly the self-energy terms due to one phonon ($n=2$ in
eq.(4)) and two phonons ($n=4$ in eq.(4)) scattering processes
which determine the renormalized electron mass $m_{eff}$ through
the relations:

\begin{equation}
{{m_{eff}}\over {m_0}}= \,{{1 - {{\partial Re\Sigma_{\bf
k}(\epsilon)}/ {\partial \epsilon }}|_{{\bf k}=0; \,\,
\epsilon=-J}} \over {1 + {{\partial Re\Sigma_{\bf k}(\epsilon)}/
{\partial \varepsilon_{\bf k} }}}|_{{\bf k}=0; \,\, \epsilon=-J}}
\label{5}
\end{equation}

where, $Re\Sigma_{\bf k}(\epsilon)=\, Re\Sigma_{\bf
k}^{(1)}(\epsilon) + Re\Sigma_{\bf k}^{(2a)}(\epsilon) +
Re\Sigma_{\bf k}^{(2b)}(\epsilon) + Re\Sigma_{\bf
k}^{(2c)}(\epsilon)$ is the frequency dependent real part of the
retarded self-energy. There are three contributions due to
different connected two-phonons diagrams \cite{mahan}. Their
effect is however confined to the intermediate regime in which
$\omega{_\pi}$ is comparable to the electronic energy $J$. We set
$J=\,0.1eV$ with the caveat that electron-electron correlations
(weak in conducting polymers with wide $\pi$-electron bands and
not taken into account by the SSH model) may become relevant in
narrow band systems. This value is lower than those usually taken
for the SSH adiabatic model but it allows us to discuss also a
broad range of (anti)adiabatic parameters with reasonable choices
of phonon energies. In the intermediate regime, the two phonons
diagrams enhance the effective mass by $\sim 15\%$ with respect to
the one phonon result. Instead, in the fully adiabatic and
antiadiabatic regimes the two phonons contributions (evaluated at
the band bottom) are negligible. Hereafter, the displayed results
depend on the very one phonon self-energy $\Sigma_{\bf
k}^{(1)}(i\epsilon_m)$ term ($\epsilon_m=(2m+1)\pi/\beta$ with $m$
integer number) whose finite temperatures analytic expression is
given by:

\begin{equation}
\Sigma_{\bf k}^{(1)}(i\epsilon_m)=\,-\sum_{\bf q} g^2({\bf k,k-q})
\Biggl[ {{n_B(\omega_{\bf q}) + n_F(-\varepsilon_{\bf k-q})}\over
{i\epsilon_m - \varepsilon_{\bf k-q} - \omega_{\bf q}}} +
{{n_B(\omega_{\bf q}) + n_F(\varepsilon_{\bf k-q})}\over
{i\epsilon_m - \varepsilon_{\bf k-q} + \omega_{\bf q}}} \Biggr]
\label{6}
\end{equation}

$n_B$ and $n_F$ are the Bose and Fermi occupation factors
respectively.

Figure 1 shows, both in one and two dimensions, a sizeable mass
enhancement in the intermediate  regime with a pronounced spike at
$\omega_{\pi} \sim \sqrt{2} J$. In 2D the effective mass is larger
than in 1D. The onset of a mass renormalization starting at
$\omega_{\pi} \sim J/2$ and, more evidently, at $\omega_{\pi} \sim
J$ (together with the increased relevance of multiphonons
contributions) signals that polaron formation is expected in this
regime while no mass enhancement is obtained in the adiabatic and
antiadiabatic limits. The same trend is observed both in 1D and
2D. Let's analyse in detail the origin of the divergent-like mass
behavior. The main contribution to both  self-energy partial
derivatives in eq.(5) comes from the lattice mode vectors
connecting two electronic states such that $J + \varepsilon_{\bf
q} - \omega_{\bf q} \sim 0$ hence, from {\bf q}-vectors satisfying
the relation $${{J \sqrt{2}} \over {\omega_{\pi}}} \sim \sqrt{1 -
cos({\bf q} \cdot {\bf a})}$$ Although for any value  $J <
\omega_{\pi}$ a set of singular {\bf q}-vectors does exist, their
divergent contributions to the numerator and denominator in eq.(5)
generally cancel out and no substantial effect is seen on the
effective mass. Only in the case $\omega_{\pi} \sim \sqrt{2} J$
something special occurs due to scattering by phonons at the
points such that $|q_x + q_y|=\,\pi/2$. Infact ${{\partial
Re\Sigma_{\bf k}(\epsilon)}/ {\partial \varepsilon_{\bf k} }}$
contains as modulation factor a $cos({\bf q} \cdot {\bf a})$ term
which is peculiar of the band structure and responsible for the
van Hove singularities in the density of states. The vanishing of
this term at the same wave vectors which allow for energy
conservation finally results in the abrupt increase of the
effective mass observed in Fig.1. In the square lattice the
divergent-like behavior is more evident since many points (only
one point in 1D) fulfill the simultaneous occurence of the two
singular effects. It should be emphasized that this phenomenon,
rather than being a general feature of two dimensional systems, is
related to the peculiarities of the electron band in the square
lattice.

Let's look now at the spectral function defined by $A({\bf k},
\,\epsilon)=\,-2Im G_{ret}({\bf k}, \,\epsilon)$ to get more
insight \cite{robin} into the suggestions proposed by the
effective mass behavior. In terms of the retarded self-energy,
obtained by eq.(6) through analytic continuation $\epsilon_m \to
\epsilon + i \delta$, $A({\bf k}, \,\epsilon)$ reads:

\begin{equation}
A({\bf k}, \,\epsilon)= \,2\pi \delta\bigl(\epsilon -
\varepsilon_{\bf k} - Re\Sigma_{\bf k}(\epsilon)\bigr) + {{(-) 2
Im \Sigma_{\bf k}(\epsilon)} \over {\Bigl(\epsilon -
\varepsilon_{\bf k} - Re\Sigma_{\bf k}(\epsilon) \Bigr)^2 + \Bigl[
Im \Sigma_{\bf k}(\epsilon) \Bigr]^2}} \label{7}
\end{equation}

The first addendum in eq.(7) contributes when $Im \Sigma_{\bf
k}(\epsilon)=\,0$. The band bottom ($|{\bf k}|=\,0$) spectral
function has been computed in a number of representative cases and
numerical convergence has been achieved by summing the self-energy
term (eq.(6)) over 6000 $q-$points in the 1D Brillouin zone and
90000 ${\bf q-}$ points in the reduced 2D Brillouin zone. The
complexity of the numerical work is mainly related to the search
of the zeros of the $\delta$-function argument in eq.(7) and in
the $Im \Sigma_{k=\,0}(\epsilon)$. In 1D, we have used the
representation $$\delta[f(q)]=\,{{\delta(q-q^0)} \over
{|df/dq|_{q=\,q^0}}}$$ and obtained convergence by summing over
$10^5$ points of the $\epsilon-$ axis. In 2D, the sum over {\bf q}
vectors can be handled as follows: $${1 \over N}\sum_{\bf q}
\rightarrow {1 \over N_x}\sum_{q_x}\sqrt{V}\int_{-\infty}^{\infty}
{{dq_y} \over {2\pi}}$$ with $V$ being the cell volume and, at a
fixed $q_x$, the $\delta$-function transforms as
$$\delta[f(q_x,q_y)]=\,{{\delta(q_y-q_y^0)} \over {|\partial
f/\partial q_y|_{q_y=\,q_y^0}}}.$$ At any $\epsilon$ (we take
80000 points) the program searches the $(q_x,q_y)$ points which
allow for energy conservation and the $Im
\Sigma_{k=\,0}(\epsilon)$ is normalized over the total number of
these pairs.

 The sum rule $$\int_{-\infty}^{\infty}{{d\epsilon}
\over {2\pi}} A(|{\bf k}|=\,0, \,\epsilon) =\,1$$ has to be
numerically fulfilled in principle. Here however we are
approximating the total self-energy by the one-phonon term
(eq.(6)) which linearly depends on the free parameter
$\alpha^2/4K$. As the {\it e-ph} coupling grows multiphonons terms
become more relevant and our approximation becomes less accurate.
Accordingly, deviations from the sum rule are expected as a
measure of the loss of spectral weight associated with higher
order self-energy effects. In this regard, the sum rule numerical
analysis permits to define the range of $\alpha^2/4K$ values
within which the one phonon approximation is reliable. For any
choice of input parameters we are able to estimate the intrinsic
error of our physical model.

Let's start the discussion looking at the spectral function in the
adiabatic regime (Figures 2). In one dimension and at very weak
coupling (Fig.2(a)) the sum rule is satisfied and the spectral
weight spreads mainly in a few peaks around the highest one
located at the energy $\epsilon =\,-100meV$ with a significant
tail up to energy levels of order $\epsilon \sim \,-80meV$. At
larger couplings (Fig.2(c)), where the dimensionless effective
coupling is $\alpha^2/(4KJ)=\,0.1$, the tail is appreciable up to
$\epsilon \sim \,-50meV$ and the loss of spectral weight is $\sim
20\%$, being $\int_{-\infty}^{\infty}{d\epsilon} A(k=\,0,
\,\epsilon)/{2\pi} =\,0.8$. A well defined transition is however
still present at $\epsilon =\,-108.2meV$. In two dimensions, a few
well resolved transitions around $\epsilon =\,-100meV$ show up at
very weak couplings (Fig.2(b)) and the sum rule is again satisfied
but the loss of spectral weight grows to $\sim 50\%$ at larger
couplings (Fig.2(d)). The breakdown of the one phonon
approximation is therefore dimension dependent and, by increasing
the {\it e-ph} coupling, 2D systems seem to favour the appearance
of multiphononic contributions in the adiabatic regime.

Figures 3 deal with the intermediate regime $\omega_{\pi}=\,J$.
Electrons are still good excitations in the extremely weak
coupling and one dimensional case (Fig. 3(a)) with one well
defined transition at $\epsilon =\,-100.9meV$ but, at larger
couplings (Fig. 3(c)), the sum rule is far from being satisfied
and a $40\%$ loss of spectral weight is observed together with a
strong reduction of the main peak height whose position is shifted
to $\epsilon =\,-119meV$. In 2D, the sum rule is fulfilled at very
weak couplings (Fig. 3(b)) but the appearance of several
transitions peaks in the range $[-110, -90]meV$ signals the onset
of a polaronic state. At larger couplings (Fig. 3(d)) the sum rule
is strongly violated and the electronic quasiparticle picture
totally breaks down. Note that the absorption spectra broaden
(both in 1D and 2D) by enhancing the strength of the {\it e-ph}
coupling as a consequence of the mixing of electronic states and
lattice vibrational excitations. Although comparisons with
specific data are not possible at this stage, broad photoemission
spectra are known \cite{mello,alexkab,timusk} to be a feature of
systems with polaronic charge carriers.

Figures 4 illustrate that electrons are good quasiparticles in the
fully antiadiabatic regime in one and also in two dimensions. In
1D, for both values of the {\it e-ph} coupling there is a well
resolved peak due to the $\delta-$function contribution and
located at $\epsilon =\,-98.8meV$ (Fig. 4(a)) and $\epsilon
=\,-90.4meV$ (Fig. 4(c)), respectively. The sum rule is well
satisfied in Fig. 4(a) while a slight loss of spectral weight
occurs in Fig. 4(c)  being $\int_{-\infty}^{\infty}{d\epsilon}
A(k=\,0, \,\epsilon)/{2\pi} =\,0.92$. In 2D the spectra do not
exhibit any relevant change with respect to the corresponding 1D
cases: in Fig.4(b) a clear transition appears at $\epsilon
=\,-98.5meV$ while the peak is located at $\epsilon =\,-89meV$ in
the moderately weak coupling case of Fig. 4(d). In the
antiadiabatic regime the main transitions are always due to the
first addendum on the r.h.s. of eq.(7). We point out that also in
the different context of the excitonic spectral function
\cite{kongeter} the disappearance of side peaks in the absorption
probability had been predicted in the antiadiabatic regime due to
the fast phonon fluctuations which destroy the high-lying excited
states in the potential well.

\section*{3. Final Remarks}

The Su-Schrieffer-Heeger tight binding model Hamiltonian has been
extended to the study of a {\it two} dimensional electron-lattice
structure. Through a perturbative approach and an exact
computation of low order diagrams, we study the renormalization of
the charge carrier effective mass versus the adiabaticity
parameter both for a linear chain and for a square lattice. In the
intermediate regime, where the phonons compete with the electrons
on the energy scale, we find a sizeable mass enhancement which may
be understood as a signature of polaron formation. This
enhancement is more pronounced in the square lattice mostly at
$\omega_{\pi} \sim \sqrt{2} J$.  The analysis of the electron
spectral function shows that the model Hamiltonian hosts quite
different behaviors according to the regime set by the adiabatic
parameter. We have computed the spectral function at the bottom of
the band in a number of representative cases by varying the
strength of the effective coupling and using the spectral function
sum rule as a testing bench for the reliability of our one phonon
approximation. While in antiadiabatic conditions the electrons
behave as good quasiparticles both in one and two dimensions,
novel features emerge in the moderately adiabatic and intermediate
regime where multiphononic terms become appreciable by increasing
the strength of the {\it e-ph} coupling, the spectral weight is
progressively spread among several transition peaks and the
electronic quasiparticle picture is lost. Unlike the Holstein
model whose ground state polaronic properties are essentially
dimension independent, we find that the onset of a polaronic state
is more likely to occur in 2D than in 1D thus confirming the trend
of the effective mass computation and suggesting that the
Su-Schrieffer-Heeger Hamiltonian is rather sensitive to
dimensionality effects.

\begin{figure}
\vspace*{1truecm} \caption{Renormalized masses (in units of bare
band electron mass) versus the adiabaticity parameter in one and
two dimensions. $m_{eff}^{(1)}$ is due to the one phonon
self-energy correction. The coupling constants are in $meV$.}
\end{figure}

\begin{figure}
\vspace*{1truecm} \caption{1D and 2D Electron spectral functions
in the adiabatic regime and (a,b) extremely weak {\it e-ph}
coupling; (c,d) moderately weak {\it e-ph} coupling. $J=\,0.1eV$.}
\end{figure}

\begin{figure}
\vspace*{1truecm} \caption{1D and 2D Electron spectral functions
in the intermediate regime and (a,b) extremely weak {\it e-ph}
coupling; (c,d) moderately weak {\it e-ph} coupling. $J=\,0.1eV$.
}
\end{figure}

\begin{figure}
\vspace*{1truecm} \caption{1D and 2D Electron spectral functions
in antiadiabatic regime and (a,b) extremely weak {\it e-ph}
coupling; (c,d) moderately weak {\it e-ph} coupling. $J=\,0.1eV$.
}
\end{figure}

\section*{Acknowledgements}

This work has been done at the Coop "Il Rafano" in Trequanda.

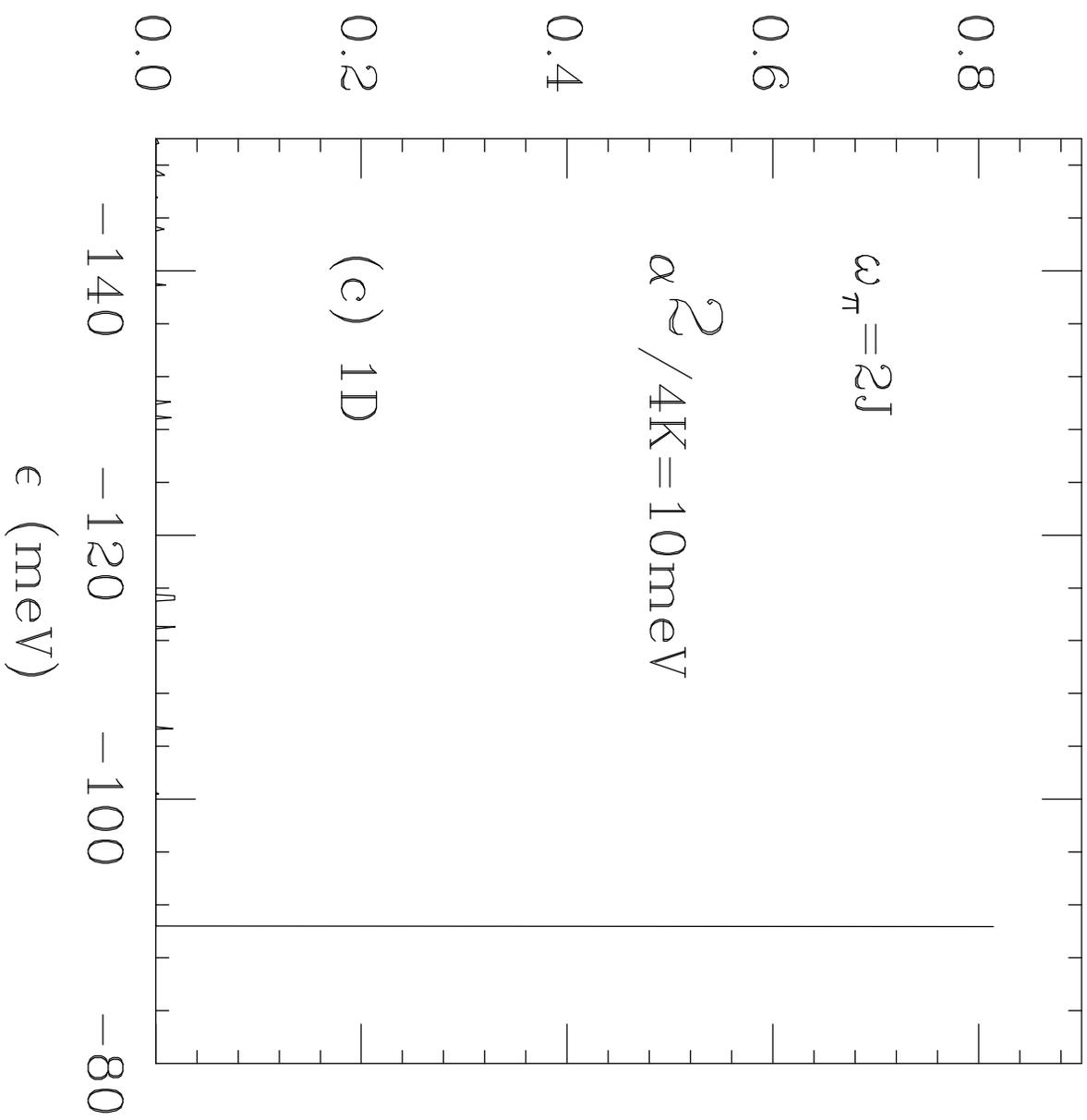
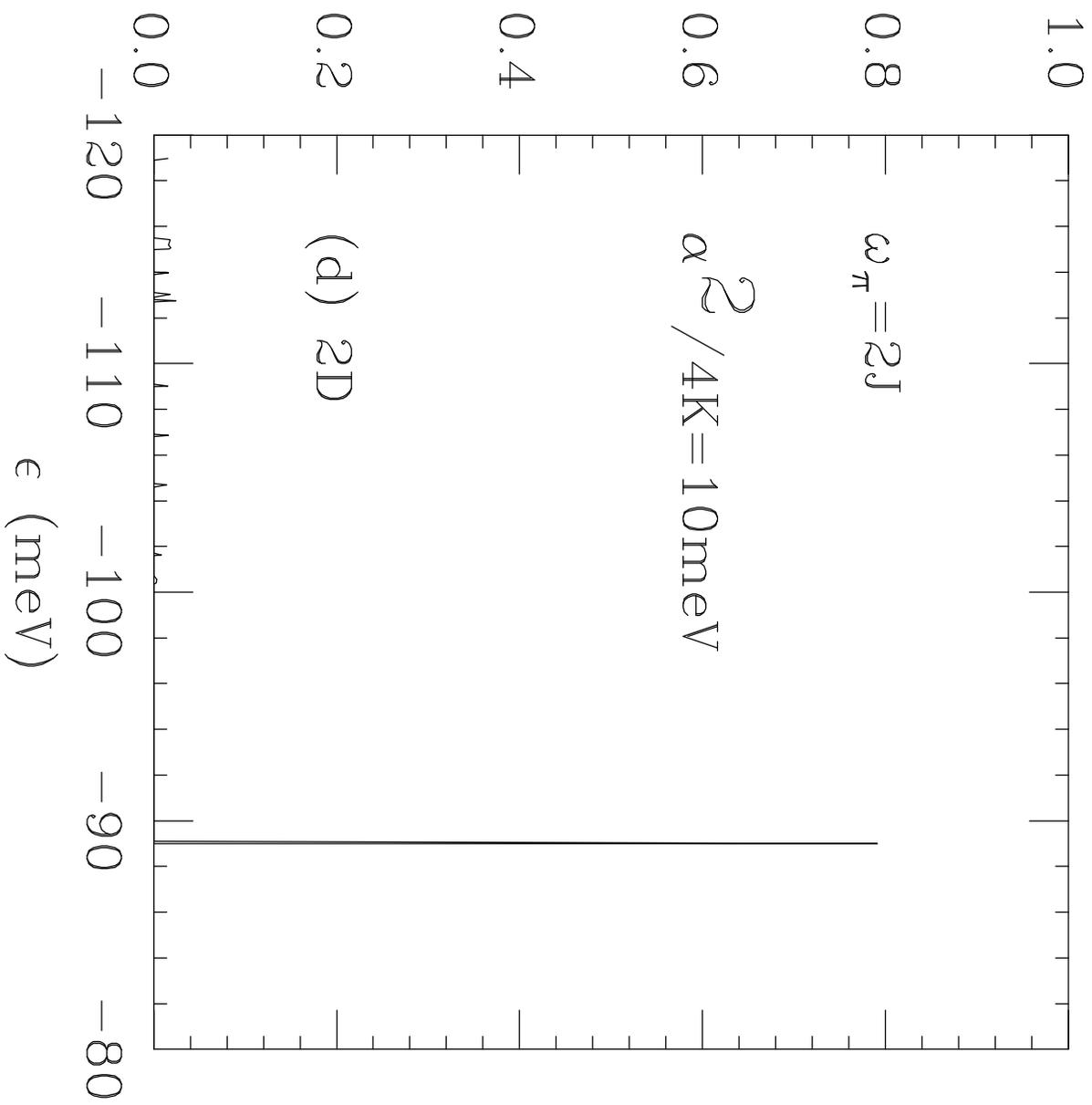
\end{document}